\documentclass{elsart4-1}

\usepackage{amssymb}
\usepackage{amsmath}
\usepackage{amsfonts}
\usepackage{amssymb}

\usepackage[english,francais]{babel}

\newtheorem{e-proposition}[theorem]{Proposition}

\newtheorem{e-definition}[theorem]{Definition\rm}

\renewcommand{\theequation}{\arabic{equation}}
\def\og{\leavevmode\raise.3ex\hbox{$\scriptscriptstyle\langle\!\langle$~}}
\def\fg{\leavevmode\raise.3ex\hbox{~$\!\scriptscriptstyle\,\rangle\!\rangle$}}

\def\be{\begin{equation}}
\def\ee{\end{equation}}

\begin{document}

\centerline{Physics/Header}
\begin{frontmatter}


\selectlanguage{english}
\title{Exact solution of the three-boson problem at vanishing energy}


\selectlanguage{english}
\author[authorlabel1]{Christophe Mora},
\ead{mora@lpa.ens.fr}
\author[authorlabel2]{Alexander O. Gogolin},
\ead{a.gogolin@ic.ac.uk}
\author[authorlabel3]{Reinhold Egger}
\ead{egger@thphy.uni-duesseldorf.de}

\address[authorlabel1]{Laboratoire Pierre Aigrain, ENS, 
Universit\'e Denis Diderot 7, CNRS; 24 rue Lhomond, 75005 Paris, France}
\address[authorlabel2]{Department of Mathematics, Imperial College
  London, 180 Queen's Gate, London SW7 2AZ, UK}
\address[authorlabel3]{Institut f\"ur Theoretische Physik, 
Heinrich-Heine-Universit\"at, D-40225 D\"usseldorf, Germany}


\medskip
\begin{center}
\end{center}

\begin{abstract}
A zero range approach is used to model resonant two-body interactions
between three identical bosons. A dimensionless phase
parametrizes the three-body boundary condition while  the
scattering length enters the Bethe-Peierls boundary condition.
The model is solved exactly at zero energy for any value of the
scattering length, positive or negative. From this solution, an
analytical expression for the rate of three-body recombination 
to the universal shallow dimer is extracted.

\vskip 0.5\baselineskip

\selectlanguage{francais}

 \vskip 0.5\baselineskip
}
\end{abstract}
\end{frontmatter}


\selectlanguage{english}
\section{Introduction}
\label{intro}

The few-body problem is a fundamental tool in the study of dilute
degenerate gases such as ultracold atomic systems~\cite{braaten2006}. It provides a basis
to construct effective low energy many-body theories~\cite{bloch2008,braaten1999,bedaque1999,braaten2002,petrov2004,mora2005b}. In addition,
two- and three-body scattering properties determine the recombination 
and loss rates~\cite{nielsen1999low,esry1999,bedaque2000,petrov2004b}
 in atomic vapors and are thus of practical 
importance. The last decade has witnessed the rapid progress in the
Feshbach resonance technique~\cite{koehler2006} 
which allows to tune experimentally the
two-body scattering length $a$. An intense experimental and theoretical
activity has then developed to investigate ultracold atoms with
resonant short-ranged interactions characterized by large scattering
lengths~\cite{braaten2002}.
Universality~\cite{braaten2006,platter2009} emerges in this regime, in
particular 
at the level of few-body physics~\cite{dincao2005}, for which details of the potential
at short distances become unessential. In this context, universal
predictions can be captured by zero range potentials where atomic
interaction enters only via boundary conditions imposed on the
wavefunctions.  

For identical bosons, Efimov~\cite{efimov1970} has predicted a
universal hierarchy of shallow bound trimer states. The 
corresponding energies form a geometric spectrum, which signals a
discrete scaling symmetry. At unitarity, there is an infinite number
of such trimers with an accumulation point at the 
continuum threshold. The first evidence of
Efimov physics by a resonant enhancement of three-body recombination
was reported in Ref.~\cite{kraemer2006} with $^{133}$Cs atoms.
This observation has been followed by many experiments on Efimov physics
involving various atoms, fermionic or
bosonic~\cite{knoop2009,zaccanti2009,barontini2009,gross2009,ottenstein2008,huckans2009,pollack2009}.
Note that the Efimov effect does not require identical bosons. For three atoms
with identical mass, operators exchanging atoms commute with the
Hamiltonian. The presence of a fully symmetric sector is then
sufficient for the Efimov effect to develop. This rules out for instance
identical fermions but it is compatible with fermions in three
different spin states.

A vast literature is already devoted to the study of the three-body problem. 
There exist recent and comprehensive
reviews~\cite{braaten2006,platter2009,braaten2007,nielsen2001} 
on the subject covering a variety of approximate 
and exact techniques developed for this problem. In a remarkable series of papers,
Gasaneo, Macek and Ovchinnikov have introduced the so-called Sturmian method which allows 
to derive some exact results in the universal, or zero range, case. They first solved a 
simple model~\cite{gasaneo2001} where only two atoms interact, and
then obtained an exact solution~\cite{macek2005} 
for the three-body problem at zero negative energy and for the atom-dimer scattering length.
Finally, they were able to extend~\cite{macek2006} the exact solution to an arbitrary positive energy.
The zero-energy limit is feasible but remains quite involved.

For negative energies $E < 0$ and a positive scattering length, the
system of three bosons only exists
as a single atom on one side, and a dimer on the other side binding 
the two remaining atoms. This atom-dimer compound is entangled at short
distance and decouples at large distance.
The situation changes dramatically when the zero-energy $E=0$ threshold 
is crossed and a new channel opens: the three atoms can also
separate freely at large distance. Therefore the zero-energy wavefunction
is singular: it depends on whether the zero-energy limit is taken from
positive or negative values.

In this paper we investigate the case of three identical bosons
interacting via zero range potential with an arbitrary scattering
length $a$. In order to proceed with a well-defined model, 
the Bethe-Peierls two-body boundary condition is
supplemented by a three-body boundary condition. 
The resulting model is solved exactly at zero
positive energy. As an application, an exact formula is derived for
the three-body recombination rate in agreement with previous works.
The approach developed in this paper has similarities with the Sturmian method
of Refs.~\cite{gasaneo2001,macek2005,macek2006}. It is nonetheless based on an integral equation
that derives from the Schr\"odinger equation, a method pioneered in Ref.~\cite{skorniakov1956}. Part of the results presented in this paper have already been
shown in a preceding letter~\cite{gogolin2008}.
The zero-range model is introduced and reviewed in Sec.~\ref{zero}. The solution
of the three-boson problem at zero energy is explicited in Sec.~\ref{solution} both
for positive and negative scattering lengths. It is used in Sec.~\ref{rate} to
derive the rate for three-body recombination. Sec.~\ref{conclusion} summarizes the
results obtained in this paper.

\section{Zero-range model}
\label{zero}

\subsection{Integral equation}

We introduce a zero-range model for three identical bosons of mass
$m$. The three-boson wavefunction is simply the solution of the free Schr\"odinger
equation with, in addition, the Bethe-Peierls boundary condition
\be\label{bethe}
\left. \psi({\bf x},{\bf y}) \right|_{y\to0} = \frac{f({\bf x})}{4 \pi}
\left(\frac{1}{y} - \frac{1}{a} \right),
\ee
which recovers the two-body physics. ${\bf y}$ is the distance between
two bosons (denoted $1$ and $2$) and $\sqrt{3} {\bf x}/2$ the distance
between their centroid and the third boson (denoted $3$), the center
of mass being decoupled.  
The two other two-body boundary conditions follow from symmetrization
of the wavefunction.
 The Schr\"odinger equation on $\psi$ can be transformed 
into an integral equation for the reduced atom-dimer function $f({\bf x})$. 
The corresponding procedure is now
standard~\cite{petrov2004,mora2005b,petrov2004b,skorniakov1956,petrov2003,mora2004,mora2005}
and we will only briefly review it here.
Applying the Laplacian on the three boundary conditions implied from Eq.~\eqref{bethe},
the Schr\"odinger equation acquires a source term 
\be\label{schro}
- \left( \nabla_{\bf x}^2 +\nabla_{\bf y}^2 + \frac{m E}{\hbar^2}
\right)  \psi({\bf x},{\bf y}) = {\cal S} ({\bf x},{\bf y})
\ee
with ${\cal S} ({\bf x},{\bf y}) = (1+ \hat{Q}) f({\bf x}) \delta( {\bf y})$.
The operator $\hat{Q} = \hat{P}_{23} +  \hat{P}_{13}$, where
$\hat{P}_{ij}$ exchanges atoms $i$ and $j$, ensures the
proper bosonic symmetry.

Using the Green's function
\be
G_{\varepsilon} ({\bf p}_1,{\bf p}_2) =  \frac{1}{p_1^2 + p_2^2 -
  \varepsilon},
\ee
the solution of Eq.~\eqref{schro} takes the form
\be\label{solgen}
\psi ({\bf x},{\bf y}) =\psi_0 ({\bf x},{\bf y}) \, \theta(E)
+ \int d^3 x' \, d^3 y' G_{m E/\hbar^2} ( {\bf x} - {\bf x}', {\bf y}
- {\bf y}')  {\cal S} ({\bf x}',{\bf y}'),
\ee
where $\psi_0$, a symmetrized combination of plane waves,
 describes the incoming wave. $\psi_0$ is solution 
of the homogeneous part of Eq.~\eqref{schro}.
$\theta(E)$ is the Heavyside function.
The integral equation is closed on the function $f$ by taking the
$y \to 0$ limit in Eq.~\eqref{solgen} with the Bethe-Peierls boundary
condition~\eqref{bethe}. This results in
\be\label{intequ}
 \left( \hat{L}_{m E/\hbar^2}  - \frac{1}{a} \right)
f ({\bf x}) = 4 \pi \psi_0 ({\bf x},0) \, \theta(E),
\ee
where the expression of the operator $\hat{L}_{\varepsilon}$ is given
in momentum space by
\be\label{opel}
\hat{L}_{\varepsilon} f({\bf k}) = \sqrt{-\varepsilon + k^2} \,\,
 f ({\bf k})
- \frac{2}{\sqrt{3} \pi^2} \int d^3 k' \frac{ f ({\bf k}')}{k^2+k'^2 +
{\bf k}\cdot {\bf k}' - 3 \varepsilon/4}.
\ee

\subsection{Three-body boundary condition}

The integral equation~\eqref{intequ} was first derived by Skorniakov
and Ter-Martirosian~\cite{skorniakov1956}, and shown later by
Danilov~\cite{danilov1961} to be ill-defined 
with a dense and unbounded spectrum at negative energy.
This unphysical prediction is related to the Thomas collapse\cite{thomas1935}, or {\it
  fall to the center} effect, at
vanishing distances between the three bosons.
It suggests that an additional boundary
condition~\cite{minlos1961,minlos1961b} is necessary if we wish 
to define a proper zero-range model for this problem.
The operator $\hat{L}_{\varepsilon}$ does not break the rotational
symmetry and thus
decouples the different angular momenta. We shall concentrate on the
rotationally invariant s-wave channel, $f ({\bf x}) = f(x)$, where the
{\it fall to the center} pathology does occur. At small
distances $x \to 0$, the inverse scattering length 
$1/a$, the energy $\varepsilon$ and the incoming wavefunction $\psi_0$
 can be neglected in Eq.\eqref{intequ}. Therefore all wavefunctions
become solutions of the universal homogeneous equation
\be \label{integral}
\hat{L}_0 f (x) = 0.
\ee
The operator $\hat{L}_0$ has a simple action on power functions (see
appendix~\ref{appen:bound}), 
\be\label{homo}
\hat{L}_0 x^{-(1+\nu)} = \phi (\nu) x^{-(2+\nu)}
\qquad \qquad 
\phi (\nu) = \frac{\nu }{\tan(\pi \nu/2)} G(\nu),
\ee
where the function
\be \label{funcG}
G (\nu) = 1 - \frac{8}{\sqrt{3}} \frac{\sin( \nu \pi /6)}{\nu \cos(\nu
  \pi /2)}
\ee
has been defined.
$G(\nu)$ has an infinite number of roots on the real axis and two
complex conjugate imaginary roots at
$\nu = \pm i s_0$ with $s_0 \simeq 1.00624$. They correspond to the
incoming and outgoing solutions to Eq.~\eqref{integral},
$\frac{e^{\pm i s_0 \ln x}}{x}$,  that dominate the
$x \to 0$ asymptotic behavior of $f$. All other solutions to Eq.~\eqref{integral} attenuate more
rapidly as $x$ tends to zero. The limiting form of $f(x)$ is thus not fully determined since
any combination of these two complex conjugate solutions is admissible.
This has been shown~\cite{danilov1961} to lead to an unphysical dense and unbounded spectrum: there exists an eigenstate
for any (negative) value of the energy $E$.

The three-body problem becomes well-defined if one imposes the
additional boundary condition~\cite{efimov1970,danilov1961,minlos1961} 
\be \label{bound3}
\left. f (x) \right|_{x\to0} \propto 
\frac{\sin[s_0 \ln (x/\ell) + \varphi]}{x},
\ee
for the asymptotic behaviour at vanishing $x$, {\it i.e.} when the three
bosons all coincide. The length $\ell$ does not bear any physical
meaning, its only purpose is to set up a reference length in the
logarithm. It is defined up to a multiplicative factor $e^{\pi/s_0}$
and can therefore take arbitrarily small or large values. 
All possible boundary conditions are exhausted when $\varphi$ is taken
in the interval $[0,\pi]$, and changing the value of $\ell$ only amounts
to shift the origin of $\varphi$. Note that
$\varphi$ has a clear physical meaning: it can be thought of as the phase shift
between the incoming and 
outgoing waves with respect to  the origin $x=0$. The amplitudes of these two
waves are equal as a result of flux conservation. This property is
however violated once the recombination to deep dimer states is included in
the formalism~\cite{nielsen2002,braaten2003}. 

Eq.~\eqref{bound3} can be given a transparent physical meaning. The
phase shift $\varphi$ is determined by short-range three-body physics
occurring at inter-particles distances for which the zero-range approach is not valid.
This short-distance physics is not sensitive to the larger
length scales of the problem - such as the scattering length $a$ - and
$\varphi$ appears as a universal phase that applies to all
eigenstates of the three-body problem.
$\varphi$ thus plays the role of a three-body parameter. It is the analogue of the
scattering length $a$ in the two-body problem. 
Finally, Eq.~\eqref{bound3} exhibits a discrete scaling symmetry $x \to x
e^{\pi/s_0}$ from which the geometrical pattern of 
Efimov states originates.

\subsection{Two-channel model}

For a model with
a finite range potential of size $\ell_0$, the {\it scaling limit}
is such that $\ell_0$ is sent to zero while the scattering length $a$
is kept finite~\cite{braaten2006}.
A zero range model is expected to describe directly the {\it scaling
  limit} of a more realistic finite range potential.
Enforcing a particular set of boundary conditions to the 
solutions of the free  Schr\"odinger equation does not automatically
constitute a well-defined zero range model. It is indeed notoriously
difficult to prove rigorously that the corresponding model is
self-adjoint. We have circumvented these difficulties in Ref.~\cite{gogolin2008}
by introducing a two-channel model: interactions between atoms in an
open channel are
mediated by a molecular state in a closed channel. In addition to the
scattering length $a$, this model introduces an additional 
 length $R^*$ that corresponds to the size of the molecular state
coupled to the open channel continuum. Similarly to
Refs.~\cite{petrov2004b,fedorov2001}, it amounts to give some energy
dependence to the scattering length $a$, and the length $R^*$ is proportional to the
s-wave {\it effective range}~\cite{braaten2006}.
In addition to providing a quantitative description of a narrow
Feshbach resonance, the two-channel model can also be seen as a toy
model describing a generic two-body interaction in the {\it scaling limit}.
It is moreover self-adjoint by construction and thus
regularizes the pathologies associated with the three-body problem.

The {\it scaling limit} $R^* \to 0$ of the two-channel model introduced in
Ref.~\cite{gogolin2008} can be checked to reproduce the boundary
conditions Eqs.~\eqref{bethe} and~\eqref{bound3}. Within the
two-channel model with $R^* \to 0$, Eq.~\eqref{intequ}
is replaced by 
\be\label{intequ3}
 \left( - R^* \nabla_{\bf x}^2 + \hat{L}_{\varepsilon}  - \frac{1}{a} \right)
f ({\bf x}) = 4 \pi \psi_0 ({\bf x},0) \, \theta(E),
\ee
where $\varepsilon = m E/\hbar^2$.
Since $R^* \ll a$, this equation can be split
into two integral equations. For $x \gg R^*$, the term~$\propto R^*$
inside the parenthesis can be neglected and $f ({\bf x})$ is
solution of Eq.~\eqref{intequ}. On the other hand, for $x \ll a,
1/\sqrt{|\varepsilon|}$, all
$a$ and $\varepsilon$ dependences disappear and Eq.~\eqref{intequ3} simplifies to
\be\label{intequ4}
 \left( - R^* \nabla_{\bf x}^2 + \hat{L}_{0}   \right)
f ({\bf x}) = 0.
\ee
The complete solutions to Eq.~\eqref{intequ3} are obtained by matching the 
asymptotes of the different solutions in the region $R^* \ll x \ll a$,
where the validity domains of Eqs.~\eqref{intequ} and~\eqref{intequ4}
overlap and Eq.~\eqref{integral} holds. In other words,
Eq.~\eqref{intequ4} serves as a boundary condition 
for the solutions of Eq.~\eqref{intequ}.
As detailed in appendix~\ref{appen:bound}, the solution to Eq.~\eqref{intequ4}
is obtained following Ref.~\cite{gogolin2008}. For $R^* \ll x \ll a$, one
recovers the asymptotic form~\eqref{bound3} with $R_0 \equiv \ell \, e^{- \varphi/s_0}
 \simeq 0.577 R^*$.

\section{Zero energy solution}
\label{solution}

Having established a well-defined zero-range approach for
three bosons, we proceed with the construction of the solution at zero
(positive) energy. We shall solve Eq.~\eqref{intequ} at $E=0^+$ 
with the boundary condition~\eqref{bound3}.

The incoming wave is generally a symmetrized combination
of plane waves, $$\psi_0 ({\bf r}_1,{\bf r}_2,{\bf r}_3) = 
\frac{1}{\sqrt{6} V^{3/2}} \sum_{\sigma} P_\sigma e^{i \sum_{i=1}^3 
{\bf k}_i \cdot {\bf r}_i}$$ where the summation includes
all permutations $P_\sigma$ of $\{ 1,2,3\}$.
$V$ is the system volume and the wavefunction has been normalized to one.
All wavevectors ${\bf k}_i$ vanish at zero energy, hence
$\psi_0 ({\bf x},0) = \sqrt{6/V^3}$. Let us define the auxiliary
function $\tilde{f}(x)$ such that
\be \label{defft}
f(x) = - 4 \pi a \sqrt{\frac{6}{V^3}}  \left( 1 - \frac{4}{\sqrt{3}}
    \frac{a}{x} + \frac{4}{\sqrt{3}} \phi (0) \tilde{f} (x) \right),
\ee
where $\phi(\nu)$ is defined Eq.~\eqref{homo}. In particular, one has
$\phi(-1) = -4/\sqrt{3}$ and $\phi(0) = \frac{2}{\pi} G(0) =
\frac{2}{\pi} [1-4\pi/(3 \sqrt{3})]$. 
The property~\eqref{homo} of the operator $\hat{L}_0$ on power 
functions yields the following integral equation 
\be \label{intequ2}
\left( \hat{L}_0  - \frac{1}{a} \right)
\tilde{f} ({\bf x}) = \frac{a}{x^2}.
\ee
An exact solution to this equation is given in the next four subsections.
Subsections~\ref{homogeneous},~\ref{asympto} and~\ref{complete} are
devoted to the case of a positive scattering length $a>0$, while 
the case $a<0$ is also exactly solved in Subsection~\ref{negative}.
An alternative way of expressing the solutions to Eq.~\eqref{intequ2} 
has been proposed in Ref.~\cite{helfrich2010}.

\subsection{Solutions of the homogeneous equation}
\label{homogeneous}

We consider the case $a>0$.
The homogeneous equation associated to Eq.~\eqref{intequ2} corresponds
to replace the right-hand-side of Eq.~\eqref{intequ2} by zero.
The solution to this integral equation can be written as a Barnes-type
contour integral
\be \label{solhomo}
\beta_1^{\pm} (x) = \frac{a}{ x} \int_{-i \infty}^{+i\infty}
\frac{d \nu}{ 2 i \pi} \, C_{\pm} (\nu) \Gamma(\nu) \sin (\pi \nu/2)
\left( \frac{x}{a} \right)^{-\nu},
\ee
with the Gamma function $\Gamma (z)$. The integration contour runs
 on the right of the imaginary axis with a small and positive real part.
$C_{\pm}$ is a set of two functions to be determined.
We apply $\hat{L}_0$ to Eq.~\eqref{solhomo}, and then use the identities $\nu
\Gamma(\nu) = \Gamma( \nu+1)$, $\cos (\pi \nu/2) = \sin [\pi
(\nu+1)/2]$ and Eq.~\eqref{homo} to obtain
\be
\hat{L}_0 \beta_1^{\pm} (x) = \frac{1}{x} \int_{-i \infty}^{+i\infty}
\frac{d \nu}{ 2 i \pi} \, G(\nu) C_{\pm} (\nu) \Gamma(\nu+1) \sin [\pi (\nu+1)/2]
\left( \frac{x}{a} \right)^{-(\nu+1)}.
\ee
Assuming the identity 
\be\label{identity}
C_{\pm} (\nu+1) = G(\nu) C_{\pm} (\nu)
\ee
and the property that $C_{\pm} (\nu)$ has no singularities (poles) within
the stripe $0 < {\rm Re}(\nu) < 2$ of the complex plane, the integral contour can be
shifted back, $\nu \to \nu -1$, with the required result
$\hat{L}_0 \beta_1^{\pm} (x) = \beta_1^{\pm} (x)/a$.
Using the Weierstrass theorem, $G(\nu)$  can be expressed in terms of
its poles $\pm b_p$, where $b_p = 2 p +1$ ($p \in \mathrm{N}$)
 and its zeros $\pm u_p$ in the complex plane, namely
\be\label{repre}
G (\nu) = \prod_{n=0}^{+\infty} \frac{\nu^2 - u_p^2}{\nu^2 - b_p^2}.
\ee
There is an infinite number of zeros on the real axis, with $u_1 =4$, $u_2 \simeq
4.6 \ldots$, and exactly two on the imaginary axis, $u_0 = \pm i s_0$.
For $p\gg 1$, we find the analytical estimate
\begin{equation}\label{esti}
u_p = b_p +  \frac{8}{\pi\sqrt{3} (2p+1)} \times 
\left\{ \begin{array}{cc}  2, & {\rm mod}(2p+1,6)=3, \\
-1, & {\rm mod}(2p+1,6)=1,5. \end{array}   \right. 
\ee
 Using the representation~\eqref{repre} and the aforementioned
 identity  $\nu
\Gamma(\nu) = \Gamma( \nu+1)$, it can be verified that the two functions
\be \label{defC}
C_{\pm} (\nu) = \prod_{p=0}^\infty \frac{\Gamma(\nu+u_p)\Gamma(1-\nu+b_p)}{
\Gamma(\nu+b_p) \Gamma(1-\nu+u_p)}
\ee
with $u_0 = \pm i s_0$ respectively, are solutions of the identity~\eqref{identity}
with no singularity in the stripe $0 < {\rm Re}(\nu) < 2$.
To be more specific, the closest singularities from the forbidden
stripe are for $\nu = 2$ and $\nu = \mp i s_0$ for the functions $C_{\pm} (\nu)$
respectively. One also verifies that the integral in
Eq.~\eqref{solhomo} is well-defined with Eq.~\eqref{defC}, {\it i.e.}
the integrand is integrable. In fact, it is physically expected that
the two functions~\eqref{defC} exhaust the set of functions that verify
the requirements: (i) integrability, (ii) absence of poles in the forbidden
stripe and, (iii) solution of the identity~\eqref{identity}. 
Remarkably, Ref.~\cite{macek2005} has introduced an alternative expression 
for the function $C_{\pm} (\nu)$. The equivalence with the definition used here 
is not straightforward to show.

To summarize our findings, we have exhibited two independent solutions
of the homogeneous equation corresponding to Eq.~\eqref{intequ2} for
$a>0$. They are given by Eq.~\eqref{solhomo} with Eq.~\eqref{defC}.

\subsection{Asymptotic behaviour at small $x$ for $\beta_1^{\pm} (x)$}
\label{asympto}
As we shall see below, the solutions $\beta_1^{\pm} (x)$ give rise to
the incoming and outgoing asymptotes $e^{\pm i s_0 \ln x}/x$ as $x\to 0$.
The proper combination of $\beta_1^{+}$ and $\beta_1^{-}$ is therefore
necessary to match the boundary condition~\eqref{bound3}.

The asymptotic behaviour for $x \ll a$ is obtained by closing the integration contour on the left
with a semi-circle at infinity in the left half plane.
 The pole at $\nu = \mp i s_0$ gives the dominant contribution with
 the result 
\be
\beta_1^{+} (x) \simeq \frac{a}{i x} \sinh \left( \frac{\pi s_0}{2}
\right) \left| {\rm Res}(C_+,-i s_0) \Gamma(-i s_0) \right| e^{i
  [ \delta_0 + s_0 \ln(x/a)]} \propto \frac{e^{ i s_0 \ln x}}{x}, 
\qquad \qquad \beta_1^{-} (x) = [
\beta_1^{+} (x) ]^*
\ee
where  $\delta_0$ is the phase of ${\rm Res}(C_+,-i s_0)
\Gamma(-i s_0)$ and ${\rm Res}(C_+,-i s_0)$ denotes the residue of $C_{+} (\nu)$ at $\nu =
- i s_0$. Using the result~\eqref{iden1} of appendix~\ref{appen:form}, the phase
$\delta_0$ can be written $\delta_0 = \pi \gamma + {\rm Arg} \Gamma(-i
s_0) \simeq 1.5875$~\cite{macek2005}.  Here $\gamma$ is 
\be \label{gamma}
\gamma = - \frac{1}{2}  -  \frac{1}{\pi} {\rm Arg} C_+ (i s_0)
 \simeq   - 0.090518155,
\ee
correcting a typo in Eq.~(17) of Ref.~\cite{gogolin2008}.
The linear combination $\beta_1 = e^{- i \delta_1} \beta_1^+ + e^{i \delta_1}
\beta_1^{-}$, with 
\be \label{delta1}
\delta_1 = \delta_0 + s_0 \ln (R_0/a) = \pi \gamma + {\rm Arg} \Gamma (-i s_0) + s_0 \ln (R_0/a),
\ee
and $R_0 \equiv \ell \, e^{- \varphi/s_0}$, recovers the three-body
boundary condition~\eqref{bound3}.

\subsection{Complete solution for $a>0$}
\label{complete}

A solution to Eq.~\eqref{intequ2} is derived using a Barnes-type
integral similar to the homogeneous case.
It reads 
\be \label{solpar}
\beta_2 (x) = {\cal B} \left( e^{-i \delta_1} \beta_2^+ (x) - e^{i \delta_1}
\beta_2^{-} (x) \right),
\ee
with
\be\label{valB}
{\cal B} = \frac{\pi}{2 i \cos \delta_1 \sqrt{-G(0)}}.
\ee
The phase $\delta_1$ is defined Eq.~\eqref{delta1}, and
\be \label{beta2}
\beta_2^{\pm} (x) = \frac{a}{ x} \int_{-i \infty}^{+i\infty}
\frac{d \nu}{ 2 i \pi} \, C_{\pm} (\nu) \Gamma(\nu) \frac{\sin (\pi
  \nu/2)}{\tan(\pi \nu)}
\left( \frac{x}{a} \right)^{-\nu},
\ee
with the functions $C_{\pm} (\nu)$ given Eq.~\eqref{defC}.
Before expliciting this solution, it can be checked that the 
boundary condition~\eqref{bound3} is reproduced in the limit $x \ll a$.
Similarly to Sec.~\ref{asympto}, the integral contour in Eq.~\eqref{beta2}
can be closed 
around the left half plane. The asymptotic form is dominated
by the closest poles located at $\nu = \pm i s_0$ and $\nu=0$.
The contribution from the  $\nu=0$ pole is easily computed. Using
the property $C_\pm (\nu) C_\pm (1-\nu) = 1$ and Eq.~\eqref{iden2}
derived in appendix~\ref{appen:form}, it is shown to cancel exactly
the  term~$\propto 1/x$ in Eq.~\eqref{defft} when $f(x)$ is expressed
in terms of $\beta_2(x)$. As a result, the asymptotic form for $f(x)$ is
solely determined by the poles at $\nu = \pm i s_0$.
Following Sec.~\ref{asympto}, the contributions of these two poles recover
the behaviour~\eqref{bound3}.

In order to verify Eq.~\eqref{solpar}, 
we apply $\hat{L}_0$ to Eq.~\eqref{beta2}. The result is 
\be 
 \hat{L}_0 \beta_2^{\pm} (x) = \frac{1}{x} \int_{-i \infty+1}^{+i\infty+1}
\frac{d \nu}{ 2 i \pi} \, C_{\pm} (\nu) \Gamma(\nu) \frac{\sin (\pi
  \nu/2)}{\tan(\pi \nu)}
\left( \frac{x}{a} \right)^{-\nu},
\ee
where the contour lies slightly on the right of the line ${\rm
  Re}(\nu) = 1$. The main difference with the homogeneous case of
Sec.~\ref{homogeneous} is that the integrand now has a pole in the stripe $0 < {\rm Re}(\nu)
< 2$. This pole is located at $\nu=1$ due to the $\tan(\pi \nu)$ in the denominator.
The integral contour can nevertheless be shifted back, $\nu \to \nu
-1$, with the result $\beta_2^{\pm} (x)/a$ and an additional
contribution that can be evaluated from the pole's residue at $\nu=1$, namely
\be \label{resulb2}
\hat{L}_0 \beta_2^{\pm} (x) = \frac{\beta_2^{\pm}}{a} +
\frac{C_{\pm}(1)}{\pi} \frac{a}{x^2},
\ee
where $C_{\pm}(1) = \pm i \sqrt{-G(0)}$ from Eq.~\eqref{iden2} in appendix~\ref{appen:form}.
Using Eq.~\eqref{resulb2}, the function $\beta_2 (x)$ from
Eq.~\eqref{solpar} is found to be solution of the integral
Eq.~\eqref{intequ2}.

To summarize, the general solution to the integral
equation~\eqref{intequ2} is  of the form $\tilde{f} (x) = \beta_2 (x) + {\cal C}
\beta_1 (x)$, where ${\cal C}$ is an arbitrary complex coefficient.

\subsection{Exact solution for $a<0$}
\label{negative}

We finally discuss the case of negative scattering length $a<0$.
In that case, the operator $\hat{L}_0 + 1/|a|$ does not have a
homogeneous solution and can be inverted. The physical meaning
of this property is that (universal) dimers do not form for $a<0$.
The atom-dimer sector is therefore absent at large distances between
the atoms. Eq.~\eqref{intequ2}, with the boundary
condition~\eqref{bound3}, possesses a unique solution, namely
\be\label{solgen2}
\beta_3 (x) = {\cal B} \left( e^{-i \delta_1} \beta_3^+ (x) - e^{i \delta_1}
\beta_3^{-} (x) \right),
\ee
with ${\cal B}$ given Eq.~\eqref{valB} and
\be
\beta_3^{\pm} (x) = - \frac{|a|}{ x} \int_{-i \infty}^{+i\infty}
\frac{d \nu}{ 2 i \pi} \, C_{\pm} (\nu) \Gamma(\nu) \frac{\sin (\pi
  \nu/2)}{\sin(\pi \nu)}
\left( \frac{x}{|a|} \right)^{-\nu}.
\ee
Similarly to Sec.~\eqref{complete},  the boundary
condition~\eqref{bound3} is verified by closing the contour around the
left hand plane and evaluating the residues at $\nu =0$ and $\nu = \pm
i s_0$. Applying  $\hat{L}_0$ to $\beta_3$ amounts to shift the
integral contour $\nu \to \nu+1$ with a sign change. A pole at $\nu =
1$ is encountered when the contour is shifted back, leading to
\be
\hat{L}_0 \beta_3^{\pm} (x) = - \frac{\beta_3^{\pm}}{|a|} -
\frac{C_{\pm}(1)}{\pi} \frac{|a|}{x^2}.
\ee
This result, inserted into Eq.~\eqref{solgen2}, shows that $\beta_3
(x)$ is solution of Eq.~\eqref{intequ2} for $a<0$.

\section{Rate of three-body recombination}
\label{rate}

Three-body recombination is a collision process in which three
incoming atoms form a (dimer) molecule and an atom.
The binding energy of the dimer is converted into kinetic energy,
and recombination thus produces losses of atoms from the trap
in ultracold atomic vapors. This effect is often important as
it can limit experimentally the lifetime of quantum gases. It also offers
a convenient experimental tool to probe resonances in the
(few)three-body problem  that are usually accompanied by increasing
recombination~\cite{dincao2006} and therefore loss rate. 
In the case of  resonant interactions (large $a>0$), the shallow dimer has
universal features and recombination can be quantitatively captured by
a zero-range model such as the one discussed in this paper.

\subsection{Formalism and result for the recombination rate}

The general solution $\tilde{f} (x) = \beta_2 (x) + {\cal C}
\beta_1 (x)$, that was derived in Sec.~\ref{complete}, can be inserted in
Eq.~\eqref{defft} to find $f(x)$, which gives access to the complete
 three-boson
wavefunction $\psi ({\bf x},{\bf y})$ {\it via} Eq.~\eqref{solgen}. The
resulting expression is cumbersome and we 
shall not write it here. Instead, the asymptotic properties of $ \psi$
can be discussed quite generally. Two relevant sectors emerge upon
considering the large distance asymptotes. Sector I: for $x \to
+\infty$ and
$y \to +\infty$, the wavefunction corresponds to three free atoms in a
symmetrized combination of plane waves. Sector II: the second sector splits
itself into three equivalent domains. Taking $x \to +\infty$
with $y$ fixed describes the atom $3$ and a dimer, formed by atoms $1$ and
 $2$, flying apart. The two remaining domains are
obtained by exchanging the single atom $3$ by $1$ or $2$.

For both sectors, there are incoming and outgoing waves whose relative
coefficients depend on the constant ${\cal C}$. In order to calculate
the three-body recombination rate, the following scattering situation is
considered: the incoming wave is formed solely by three free atoms and the flux
probability to leak into the three domains of the atom-dimer sector is computed.
The constant ${\cal C}$ is therefore chosen in order to cancel
the atom-dimer incoming wave.

We use the notation $r = \sqrt{3} x/2$ for the distance between the atom
$3$ and the $1$-$2$ centroid. For $x \to +\infty$
with $y$ fixed, the asymptotical form $\psi({\bf x},{\bf y})
\simeq \phi_0 (y) \phi_{\rm ad} (r)$ is expected, where $\phi_0 (y) =
e^{-y/a}/(y \sqrt{2 \pi a})$ is the normalized two-body bound state
(dimer) wavefunction, and 
\be\label{phiad}
 \phi_{\rm ad} (r) = {\cal A} \sqrt{\frac{a}{8 \pi}} \frac{\sqrt{3}}{2} \frac{e^{i \frac{2}{\sqrt{3}} r/a}}{r},
\ee
where ${\cal A}$ is a constant that will be extracted from the exact
solution. The wavefunction $\phi_{\rm ad} (r)$ describes the
atom-dimer relative motion corresponding to an outgoing scattered wave
by the recombination process. 
The $y \to 0$ limit in the asymptotical form of $\psi({\bf x},{\bf y})$ can be compared
to the Bethe-Peierls boundary condition~\eqref{bethe} to extract $f$
with the result
\be\label{expf}
f (x) \simeq {\cal A} \frac{e^{i x/a}}{x},
\ee
in agreement with the behaviour of $f(x)$ at large $x$ that shall be
derived in Sec.~\ref{asymp}. Interestingly, this result shows that
the  large $x$ study of $f(x)$  is sufficient to determine the
wavefunction in the atom-dimer sector through the knowledge of ${\cal A}$.
A similar reasoning also indicates that a vanishing atom-dimer incoming wave
is equivalent to a vanishing $e^{-i x/a}/x$ term in $f(x)$ at large $x$.

Eq.~\eqref{phiad} can be used to compute the rate of three-body
recombination. The current probability associated to $\phi_{\rm ad}
(r)$ is given by $j(r) = \frac{\hbar}{2 i (2 m /3)} \left[ \phi_{\rm ad}^* (r) 
\partial_r \phi_{\rm ad} (r) - \partial_r \phi_{\rm ad}^* (r) 
 \phi_{\rm ad} (r) \right]$ where $2 m /3$ is
the atom-dimer reduced mass. The probability flux $\Phi_i$
to leave the scattering region in one of the atom-dimer domains ($i$ denotes
the indices of the single atom) can be obtained by integrating $j(r)$
over a sphere with an infinitely large radius. An additional
integration over the free center of mass position 
multiplies by the volume $V$ and one finds
\be
\frac{\Phi}{3} = \Phi_i = V \lim_{r \to +\infty} 4 \pi r^2 j(r) 
= \frac{3 \sqrt{3}}{8} \frac{\hbar \, V}{m} |{\cal A}|^2,
\ee
where $\Phi$ denotes the total flux to the atom-dimer sector, or
sector II.
It is also possible to relate this flux to the large $x$ form of the auxiliary
function $\tilde{f}$ defined Eq.~\eqref{defft}, namely
\begin{subequations}
\begin{align}
\label{asymptoft}
\tilde{f} (x) & = \tilde{\cal A} \, \frac{e^{i x/a}}{x}, \\[1mm]
\Phi_i & = 192 \sqrt{3} \pi^2 a^2 \phi^2(0)  \frac{\hbar}{m} \frac{|\tilde{\cal A}|^2}{V^2},
\end{align}
\end{subequations}
where $\phi(\nu)$ is defined Eq.~\eqref{homo} and  $\phi(0) = \frac{2}{\pi} G(0) =
\frac{2}{\pi} [1-4\pi/(3 \sqrt{3})]$.
The value of $\tilde{\cal A}$ can be extracted  from the exact solution of Eq.~\eqref{intequ2}
derived in Sec.~\ref{complete}. This calculation is postponed to Sec.~\ref{asymp}.

For a dilute gas of $N$ bosons with density $n=N/V$, the three-body 
recombination rate $\alpha_{\rm rec}$ is defined~\cite{braaten2006} such that the number of 
recombination events per time and per volume is $\alpha_{\rm rec} n^3$.
Hence,
\be
\alpha_{\rm rec} \left(\frac{N}{V} \right)^3 = \frac{1}{V} \frac{d N_{\rm rec-events}}{d t} = \frac{\Phi}{V} \frac{N (N-1)(N-2)}{6}
\simeq \frac{\Phi}{V} \frac{N^3}{6}
\ee
where $\frac{N (N-1)(N-2)}{6}$ is the number of triplets among the $N$ bosons.
The resulting expression for the  three-body recombination rate is 
\be\label{formrec}
\alpha_{\rm rec} = 96 \sqrt{3} \pi^2 a^2 \phi^2(0)  \frac{\hbar}{m} |\tilde{\cal A}|^2.
\ee

\subsection{Wavefunction at large $x$}
\label{asymp}

For $x \gg a$, the integrands in the expressions of $\beta_1 (x)$ and
$\beta_2 (x)$, Eqs.~\eqref{solhomo} and~\eqref{solpar}, develop rapid
oscillations. The asymptotical behaviour can thus be deduced
from a saddle-point analysis that is detailed in appendix~\ref{saddle}.
Focusing on the exact solution,
\be
\tilde{f} (x) = \beta_2 (x) + {\cal C} \beta_1 (x) = 
{\cal B} \left( e^{-i \delta_1} \beta_2^+ (x) - e^{i \delta_1}
\beta_2^{-} (x) \right) + {\cal C} \left( e^{-i \delta_1} \beta_1^+ (x) + e^{i \delta_1}
\beta_1^{-} (x) \right),
\ee
the incoming atom-dimer wave ($\propto e^{- i x/a}/x$) is seen to vanish with the choice
\be
{\cal C} = \frac{{\cal B}}{i} \, \frac{1 - e^{-2 \pi s_0} e^{-2 i \delta_1}}{1 + e^{-2 \pi s_0} e^{-2 i \delta_1}}.
\ee
Inserting this result  back into the large $x \gg a$ expression of $\tilde{f} (x)$, Eq.~\eqref{asymptoft} is recovered with
the coefficient
\be
\tilde{\cal A} = {\cal B} e^{i(\delta_\infty - \delta_1)} \frac{a \, e^{\pi s_0}}{2}
\left( 1 - e^{-2 \pi s_0} e^{2 i \delta_1} \right)
\left[ 1 - i \frac{{\cal C}}{{\cal B}} \, 
\frac{1- e^{-2 \pi s_0} e^{2 i \delta_1}}{1 + e^{-2 \pi s_0} e^{2 i \delta_1}} \right].
\ee
Some simple algebraic manipulations further lead to the expression
\be\label{inter}
\tilde{\cal A} = {\cal B} e^{i(\delta_\infty - \delta_1)} \frac{4 a }{i}
\frac{\sin \delta_1 \, \cos \delta_1}{ e^{\pi s_0} + e^{- \pi s_0} e^{-2 i \delta_1}}.
\ee
The identity
\be
\frac{1- e^{-2 i \delta_1}}{1+ e^{-2 \pi s_0} e^{-2 i \delta_1}} = \frac{1-e^{-2 i \delta_r}}{1 - e^{-2 \pi s_0}},
\ee
with the phase $\delta_r = \delta_1 - {\rm Arg} ( 1 +e^{-2 \pi s_0} e^{2 i \delta_1})$ is used
together with the value of ${\cal B}$ from Eq.~\eqref{valB} to derive 
\be
\tilde{\cal A} = - \frac{\pi\, a}{\sqrt{-G(0)}}  e^{i(\delta_\infty - \delta_r)} \frac{\sin \delta_r}{\sinh (\pi s_0)}.
\ee
The rate for three-body recombination is finally given by
\begin{equation}\label{recomb}
\alpha_{\rm rec} = \frac{128 \pi^2 \, (4 \pi -3 \sqrt{3})}{\sinh^2(\pi s_0)} \,
\frac{\hbar^2 a^4}{m} \sin^2 (\delta_r).
\end{equation}
This function shows periodic  oscillations as a function of $\ln (R_0/a)$, with a period given by
the scaling factor $e^{\pi s_0}$, and a dimensionless 
amplitude  $128 \pi^2 \, (4 \pi -3 \sqrt{3})/\sinh^2 (\pi s_0) \simeq 67.1177$.
In practice, $e^{-2 \pi s_0} \simeq 0.0012$ is a very small number such that $\delta_1 \simeq \delta_r$ and
the oscillations are almost  sinusoidal.

Instead of Eq.~\eqref{recomb}, an alternative formula can be obtained for the  three-body recombination rate
by taking directly the absolute value squared of Eq.~\eqref{inter} and using that
$| e^{\pi s_0} + e^{- \pi s_0} e^{-2 i \delta_1} |^2 = 4 \sinh^2 (\pi s_0) + 4 \cos^2(\delta_1)$.
The result reads
\be\label{recomb2}
\alpha_{\rm rec} = \frac{128 \pi^2 \, (4 \pi -3 \sqrt{3})}{\sinh^2 (\pi s_0)} \,
\frac{\hbar^2 a^4}{m} \frac{\sin^2 (\delta_1) \sinh^2 (\pi s_0)}{ \sinh^2 (\pi s_0)+ \cos^2 (\delta_1)},
\ee
which coincides with Eq.~\eqref{recomb} and recovers the result of Refs.~\cite{macek2006,gogolin2008,petrovunp,helfrich2010}.
Ref.~\cite{platter2009} has pointed out that the
expression~\eqref{recomb}  also derives from the use of 
the optical theorem and the knowledge of the elastic boson-diboson
scattering solution at zero (negative) energy  
determined in Ref.~\cite{macek2005}.

\section{Conclusion}
\label{conclusion}

We have solved exactly the wavefunction at vanishing energy of three
identical bosons with zero range interactions. 
The zero range model requires a set of two boundary conditions.
The first one is the standard Bethe-Peierls condition when two atoms
meet. An additional three-body boundary condition when the three atoms coalesce
is then necessary. It is  due to the {\it
  fall to the center} effect, also responsible for the emergence of
the Efimov bound states.
It is characterized by a dimensionless parameter $\varphi$ which
describes the phase shift between the outgoing and the incoming waves
toward the region where the three atoms coincide.
Close to this region, the wavefunction displays oscillation as a
function of $\ln (R/a)$ where $R$ denotes the hyperradius of the three
bosons. These logarithmic oscillations thus appear in most
observables and in particular in the three-body recombination rate.

The derivation of the exact solution is facilitated by the fact that
the operator $\hat{L}_\varepsilon$ does not involve any energy scale
at zero energy: it is homogeneous. As a result, its action is stable
in the space of power functions. The trick is then to write the exact
solution as a Barnes-type contour integral over a power function 
where the power is the integration variable. Acting with $\hat{L}_0$
simply amounts to shift the integration contour. The integrand has to
be chosen such that no pole is crossed when the contour is shifted
back to its original position.

The solution is unique in the case of a negative scattering length
where no dimer can be formed.
This is in contrast with the case of a positive scattering length
where the solution is not unique but can be parametrized 
by a single parameter. It depends on the balance between the
atom-dimer and the three-particle sectors in the incoming wave.
The three-body recombination rate is evaluated by canceling the
incoming atom-dimer wave and then by calculating the prefactor of the
outgoing atom-dimer wave. The result is an exact analytical expression
that exhibits the expected logarithmic oscillations.

 \appendix

\section{Boundary condition from the two-channel model}
\label{appen:bound}

It is more convenient to solve Eq.~\eqref{intequ4} in Fourier space,
or
\be \label{intequ5}
\left( R^* k^2 + \hat{L}_{0} \right) f ({\bf k}) = 0.
\ee
The corresponding operator is rotationally invariant and thus decouples the 
different partial waves.  For all angular momenta $l > 0$, the limit 
$R^* \to 0$ can be taken directly in Eq.~\eqref{intequ5}, and coincides
with the zero range approach.
The s-wave sector $l=0$, with $f ({\bf k}) = f(k)$, requires a particular
treatment in the {\it scaling limit} $R^* \to 0$.
One notices that the operator $\hat{L}_{0}$ is homogeneous in
the sense that it does not involve any scale. It implies that $\hat{L}_{0}$ acts 
simply on power functions, namely 
\be\label{proper}
\hat{L}_{0} k^{-\nu-2} = G (\nu)  k^{-\nu-1},
\ee
where $G(\nu)$ is given by Eq.~\eqref{funcG}. This result is obtained
with the change of variable $k' = k \, e^\xi$ in Eq.~\eqref{opel} and
the integral
\be
\int_{-\infty}^{+\infty} \frac{d \xi}{2 \pi} e^{\nu \xi}
\left( \frac{e^{2 \xi}+e^{\xi}+1}{e^{2 \xi}-e^{\xi}+1}\right)
= \frac{1}{\nu} \frac{\sin( \nu \pi /6)}{\cos(\nu  \pi /2)}.
\ee
Using the property Eq.~\eqref{proper}, one verifies that the Barnes-type
contour integral (on the right of the imaginary axis) 
\be\label{fk}
f( k) =  \int_{-i \infty}^{+i \infty}
\frac{d \nu}{ 2 i \pi} \, C (\nu) 
\left( k R^* \right)^{-\nu-2},
\ee
is solution of Eq.~\eqref{intequ5}. The function
\be
C (\nu) = \frac{\pi}{\sin[ \pi (\nu - i s_0)]} C_+ (\nu)
= \frac{\pi}{\sin[ \pi (\nu + i s_0)]} C_- (\nu),
\ee
such that $G (\nu) C (\nu) = - C(\nu+1)$,
has no poles  within
the stripe $0 < {\rm Re}(\nu) < 2$ of the complex plane.
Applying $\hat{L}_{0}$ on Eq.~\eqref{fk} and shifting the contour
as $\nu \to \nu -1$ indeed recovers Eq.~\eqref{intequ5}.

The large $x \gg R^*$ behaviour of $f(x)$ is obtained from the
asymptote of $f(k)$ at small $k R^* \ll 1$.
By closing the integration contour around the left half plane in
Eq.~\eqref{fk}, $f(k)$ is evaluated as a sum over terms with powers
at least higher than $k^{-2}$. The poles at $\nu = \pm i s_0$
give the dominant contribution at small $k$. After Fourier transform,
the result reads
\be
f (x)  \propto \frac{\sin[s_0 \ln (x/R_0) ]}{x},
\ee
with $R_0 = R^* \exp \left[ ( \pi (\gamma+1/2) + {\rm Arg} \Gamma (i
s_0))/s_0 \right] \simeq 0.577 R^*$,  where $\gamma$ is given 
Eq.~\eqref{gamma}.

 \section{Useful formulas}
 \label{appen:form}

We here list some useful relations.

First, the residue of  $C_{+}$ at $\nu = - i s_0$ is given by
\be
{\rm Res}(C_+,-i s_0) = \frac{\Gamma(2+i s_0)}{\Gamma(1-i s_0)
  \Gamma(1+ 2 i s_0)} \prod_{p=1}^\infty \frac{\Gamma(-i
  s_0+u_p)\Gamma(1+i s_0 +b_p)}{
\Gamma(-i s_0 +b_p) \Gamma(1+i s_0+u_p)},
\ee
where the $p=0$ terms have been singled out from the convergent infinite product.
The general identity $\Gamma (z) \Gamma (1- z) = \pi/\sin( \pi z)$ is
used at $z= -2 i s_0$ to obtain
\be\label{iden1}
{\rm Res}(C_+,-i s_0) = \frac{\sinh(2 s_0 \pi)}{i \pi} \, [ C_{+} (i
s_0) ]^*,
\ee
with the help of the property $\Gamma(z^*) = [\Gamma(z)]^*$.

Second, using the definition~\eqref{defC} of the functions $C_{\pm}$, 
we can write
\be
C_{\pm}(1) = \prod_{p=0}^\infty \frac{\Gamma(1+u_p)\Gamma(b_p)}{
\Gamma(1+b_p) \Gamma(u_p)} = \pm i s_0 \prod_{p=1} \frac{u_p}{b_p}.
\ee
On the other hand, the product representation Eq.~\eqref{repre} of
$G(\nu)$ can be evaluated at $\nu=0$ leading to
\be
G (0) = 1-\frac{4\pi}{3 \sqrt{3}} = -  \left( s_0\,\prod_{p=1}
  \frac{u_p}{b_p} \right)^2,
\ee
with $1-4\pi/(3 \sqrt{3}) \simeq - 1.41839$. As a result, the identity
\be \label{iden2}
C_{\pm}(1) = \pm i \sqrt{-G(0)},
\ee 
is obtained.

\section{Saddle-point analysis at large $x$}
\label{saddle}

The integral expressions~\eqref{solhomo} and~\eqref{solpar}, for $\beta_1^{\pm}$ and $\beta_2^{\pm}$,
can be evaluated at large $x \gg a$. In contrast with the small $x$ case, the integration contour 
can not be closed on the right half plane. Nevertheless, the integrand exhibits rapid 
oscillations along the imaginary axis as $x$ is increased. A saddle-point, or stationary phase, approximation
is thus carried out to extract the asymptotical behaviour.

For large $\nu \gg 1$, we have $\ln \Gamma(\nu) \simeq \ln \sqrt{2
  \pi} + (\nu -1/2) \ln \nu  - \nu$, and the coefficient $C_+ (\nu)$
defined in Eq.~\eqref{defC} becomes 
\be
\ln C_+ (\nu) \simeq  \left[ \Psi(\nu) - \Psi(1-\nu) \right] \sum_{p=0}^{+\infty} (u_p - b_p),
\ee 
with the digamma function $\Psi (\nu) = \Gamma'(\nu)/\Gamma(\nu)$.
Equation \eqref{esti} can be used to check that the summation over $p$ indeed
converges. Since $\Psi(\nu) \simeq \ln \nu$ for $\nu \gg 1$,
the limits
\be
\lim_{\nu \to \pm i \infty} C_+ (\nu) = e^{\mp (i \delta_\infty + \pi s_0)}
\ee
are obtained with the phase $\delta_\infty = - \pi \sum_{p=1}^{+\infty} (u_p - b_p) + \pi \simeq 1.736$ in
agreement with Ref.~\cite{macek2005}. Noting that $\sin (\pi \nu/2) \simeq \pm i e^{\mp i \pi \nu/2}/2$
for $\nu \to \pm i \infty$, we rescale the integral with $\nu = x z$ and approximate 
\be\label{expbeta}
\beta_1^+ (x) \simeq \frac{a}{2 \sqrt{2 \pi x}} \int_{-i \infty}^{+i\infty} \frac{d z \, \varepsilon_z}{\sqrt{z}}
e^{-\varepsilon_z (i \delta_\infty + \pi s_0)} e^{x g(z)}.
\ee
We have defined $\varepsilon_z = {\rm sgn \, (Im} z)$ and the function
\be
g (z) = z \left[ \ln z - 1 +\ln a - \frac{i \varepsilon_z \pi}{2} \right].
\ee
The integral~\eqref{expbeta} is dominated by two symmetric saddle-points on the imaginary axis,
$z_0^{\pm} = \pm i/a$, derived from the condition $g'(z_0) =
0$. Moreover, one has $g(z_0^{\pm}) = \mp i /a$ and
$g''(z_0^{\pm}) = \mp i a$. After integration over the quadratic fluctuations around the two saddle-points, 
the asymptotic form
\be\label{asympb1}
\beta_1^+ (x) \simeq -i \frac{a}{2} \left( e^{\pi s_0} e^{i \delta_\infty} \frac{e^{i x/a}}{x} 
- e^{-\pi s_0} e^{-i \delta_\infty} \frac{e^{-i x/a}}{x} \right)
\ee
is obtained for $x \gg a$. The same result is derived for $\beta_1^- (x)$ with $s_0$ replaced by $-s_0$.
In this asymptotic region $x \gg a$, Eq.~\eqref{asympb1} describes a combination of incoming and outgoing atom-dimer waves.

The calculation is similar for the case of $\beta_2^{\pm}$. The difference is expressed by the following
limit
\be
\frac{\sin (\pi
  \nu/2)}{\tan(\pi \nu)} \simeq \frac{e^{\mp i \pi \nu/2}}{2}
\ee
for $\nu \to \pm i \infty$. The final result reads
\be
\beta_2^+ (x) \simeq \frac{a}{2} \left( e^{\pi s_0} e^{i \delta_\infty} \frac{e^{i x/a}}{x} 
+ e^{-\pi s_0} e^{-i \delta_\infty} \frac{e^{-i x/a}}{x} \right),
\ee
with $s_0 \to -s_0$ for $\beta_2^- (x)$. Again incoming and outgoing waves coexist in this asymptotic form.


\bibliography{bibliographie}
\bibliographystyle{elsart-num}




\end{document}